# Mechanochemical synthesis of pseudobinary Ti-V hydrides and their conversion reaction with Li and Na

Fermin Cuevas[*], Barbara Laïk, Junxian Zhang, Mickaël Mateos, Jean-Pierre Pereira-Ramos, Michel Latroche[†]

Univ Paris-Est Creteil, CNRS, ICMPE (UMR 7182), 2 rue Henri Dunant, F-94320 Thiais, France

Fermin Cuevas: 0000-0002-9055-5880 (ORCID)

Barbara Laïk: 0000-0001-5194-4210 (ORCID)

Junxian Zhang, 0000-0003-4978-3870 (ORCID)

Mickael Mateos: 0000-0002-1443-3910 (ORCID)

Jean-Pierre Pereira-Ramos: 0000-0001-5381-900X (ORCID)

Michel Latroche: 0000-0002-8677-8280 (ORCID)

**Abstract**

Lithium-ion batteries (LiBs) based on insertion electrodes reach intrinsic capacity limits. Performance improvements and cost reduction require alternative reaction mechanisms and novel battery chemistries such as conversion reactions and sodium-ion batteries (NaBs), respectively. We here study the formation of $Ti_{1-x}V_xH_2$ hydrides ($0 \leq x \leq 1$) and their electrochemical properties as anodes in LiBs and NaBs half-cells. Hydrides were synthesized by mechanochemistry of the metal powders under hydrogen atmosphere ($P_{H_2} \sim 8$ MPa). For V contents below 80 at.% ($x < 0.8$), single-phase pseudobinary dihydride compounds $Ti_{1-x}V_xH_2$ are formed. They crystallize in the fluorite-type structure and are highly nanostructured (crystallite size $\leq 10$ nm). Their lattice parameter decreases linearly with the V content leading to hydride destabilization. Electrochemical studies were first carried out in Li-ion half cells with full conversion between $Ti_{1-x}V_xH_2$ hydrides and lithium. The potential of the conversion reaction can be gradually tuned with the vanadium content due to its destabilization effect. Furthermore, different paths for the conversion reaction are observed for Ti-rich ($x \leq 0.25$) and V-rich ($x \geq 0.7$) alloys. Na-ion half-cell measurements prove the reactivity between $(V,Ti)H_2$ hydrides and sodium, albeit with significant kinetic limitations.

---

[*] Corresponding author. E-mail : fermin.cuevas@cnrs.fr
[†] Deceased author

## 1. Introduction

Lithium-ion batteries (LiBs) are now widely used as energy storage systems for many mobile applications. However, the insertion materials currently developed for this technology reach their intrinsic capacity limits [1,2]. For instance, the theoretical reversible capacity of the graphite anode is limited to 372 mAh g$^{-1}$ and this material will become critical as poor effort has been dedicated to its recycling [3]. Performance improvements and cost reduction are expected by employing new reaction mechanisms such as conversion reactions using easily recyclable materials like metals and novel battery chemistries such as sodium-ion batteries (NaBs)[4–6].

Several families of hydride compounds reacting through a conversion mechanism have been proposed as anodes of LiBs [7–10]. They react with lithium according to the general conversion reaction $M\text{H}_y + y\text{Li}^+ + ye^- \rightarrow M + y\text{LiH}$, where $M\text{H}_y$ can be a binary, a ternary or a complex hydride, *e.g.* $MgH_2$, $LaNi_5H_6$ or $Li_3AlH_6$, respectively. They provide high specific and volumetric capacities reaching for instance 2036 mAh g$^{-1}$ and 2885 mAh cm$^{-3}$ for $MgH_2$ and 1074 mAh g$^{-1}$ and 4040 mAh cm$^{-3}$ for $TiH_2$, respectively. The reversibility of the conversion reaction, high-rate capability and long cycle-life remain challenging, though great progress has been recently achieved by using nanostructured hydrides in all-solid-state devices operating at 120°C [11–13]. Mechanochemistry under hydrogen gas is proved to be an efficient technique for the synthesis of nanostructured hydrides [14,15].

Hydrides studied in LiBs are restricted so far to well-defined stoichiometric compounds as concerns the metallic counterpart. Consequently, plateau potentials are fixed to a constant value that is dictated by the thermodynamic stability of the hydride. For instance, the conversion potentials of $MgH_2$ and $TiH_2$ are 0.56 V and 0.13 V vs. Li$^+$/Li at room temperature (RT), respectively [16,17]. Though, certain hydride forming compounds, such as V-based bcc alloys [18,19], offer the property to tune their thermodynamic hydrogenation properties by taking advantage of their solid solution behavior. Therefore, a tunable conversion potential of hydride forming bcc-alloys in LiBs is expected as a function of their composition.

In this work, we study the mechanochemical synthesis and thermodynamic stability of nanostructured $Ti_{1-x}V_xH_2$ hydrides ($0 \leq x \leq 1$) as well as their electrochemical properties as conversion anodes in LiBs. Moreover, taking advantage of the tuning of the anode potential offered by the solid solution between Ti and V atoms, we propose and confirm that $Ti_{0.3}V_{0.7}H_2$ compound can undergo conversion reaction in Na-ion half cells though poorly reversible at room temperature. The use of hydrides as conversion anodes in NaBs remains almost unexplored in the literature [20,21].

## 2. Material and methods

$Ti_{1-x}V_x$ hydrides ($x$ = 0, 0.25, 0.5, 0.7, 0.8, 0.9 and 1) were elaborated by ball-milling of Ti (99.9% purity, < 44 µm, Aldrich) and V (99.5%, < 44 µm, Stream Chemicals) powder mixtures under hydrogen gas (Alphagaz; 6N). The total metal mass was typically of 3 g for metal powder mixtures. Powder mixture and stainless-steel balls (15 mm diameter), with a ball to powder weight ratio of 60:1, were loaded in an Evico-magnetics hardened stainless-steel vial (Germany) equipped with pressure and temperature sensors. Metal loading was performed in an argon filled and purified glove box (Jacomex, < 4 ppm $O_2$, < 0.2 ppm $H_2O$). Next, the vial was evacuated under primary vacuum and the initial hydrogen pressure was fixed to *ca.* $P_{H2}$ = 8 MPa. Mechanical milling was performed in a planetary mill (Pulverisette Fristch P4) at 400 rpm with four sequences of 30 min milling and 2 h rest. Total hydrogen uptake and hydrogen absorption curves on milling were determined as described in Ref. [22]. The Ti:V metal ratio in the as-milled samples was determined by inductively coupled plasma optical emission spectrometry (ICP-EOS, Variant VistaPro), after prior dissolution in $HNO_3$ 5 wt.% aqueous solution.

Structural properties were characterized by powder X-ray Diffraction (XRD) with a D8 Advance Bruker diffractometer equipped with Cu-K$\alpha$ radiation ($\lambda$ = 0.15418 nm). To avoid air contamination, all XRD measurements were performed under argon atmosphere with the help of an airtight and X-ray transparent cap sample-holder (Brucker). Diffraction patterns were analysed by the Loopstra and Rietveld method using Fullprof software [23]. Diffraction lines were modelled using the Thompson–Cox–Hastings pseudo–Voigt profile shape function to determine crystal size and micro strain as described in Ref. [24].

The electrochemical properties of $Ti_{1-x}V_x$ hydrides, both in Li-ion and Na-ion configuration, have been measured in coin type CR2032 half-cells using BioLogic VMP3 Potentiostat. Working electrodes were prepared by mixing in an agate mortar as-milled $Ti_{1-x}V_x$ hydride, carbon Super P (99% metal basis, Alfa Aesar) and sodium carboxymethylcellulose (CMC, Mw ~250,000, Aldrich) in weight ratio 1:1:1. The mixture was pressed over Ni foam at 10 T during 5 min with an active material loading of 4 mg cm$^{-2}$. Either Li or Na disks, 16 mm in diameter and 0.5 mm in thickness, were used as counter electrodes. Three Whatman® glass fibre filters (grade GF/C, General Electrics) imbued with liquid electrolyte were used as separators. For Li-ion half cells, 100 µl of 1 M solution of lithium perchlorate (LiClO$_4$) in propylene carbonate (PC) was used as electrolyte. The working electrodes were cycled in galvanostatic mode at RT at C/40 rate (2 mol of electrons in 40h per mol of dihydride) within the potential window 0.02 – 2 V versus Li$^+$/Li. In addition, o*perando* XRD measurements during galvanostatic lithiation were performed at the same rate using a special electrochemical cell equipped with Be window developed by Leriche *et al* [25]. The cut-off potential was fixed to 0.05 V vs. Li$^+$/Li. The working electrode was a composite mixture of as milled Ti$_{0.3}$V$_{0.7}$H$_2$ hydride, CMC and Carbon SP in weight ratio 1:1:1. No Ni foam was used as current collector to avoid Ni-diffraction lines. XRD data was collected using a Bruker DaVinci diffractometer using Cu-K$_\alpha$ radiation and equipped with a Lynxeye detector. Acquisition time for each pattern was 2 h. Cycling voltammetry measurements were performed as well at scan rate of 10 µV s$^{-1}$ within the same voltage range.

Most of the synthetised Ti$_{1-x}$V$_x$ hydrides were tested in Li-ion configuration while only Ti$_{0.3}$V$_{0.7}$H$_2$ hydride was studied in Na-ion half cells, using NaClO$_4$ in PC as electrolyte. The working electrode was cycled in galvanosatic mode at RT at a C/40 rate within the potential window 0 – 1.5 V versus Na$^+$/Na. Moreover, forced sodiation experiments were performed by the galvanostatic intermittent titration technique (GITT) method with alternating sodiation steps at C/40 rate with a cut-off potential of 0 V and relaxation periods for 10h at open circuit potential.

### 3. Results
### 3.1 Hydride synthesis and characterisation

Figure 1 shows the hydrogen uptake curves of Ti$_{1-x}$V$_x$ powder mixtures. Except for pure Vanadium, $x = 1$, hydrogen content keeps stable after milling over 15 min. For $0 \leq x \leq 0.8$, the hydrogen content stabilizes between 1.6 and 1.8 hydrogen atoms per formula unit (H/f.u), while for the V-rich mixture, $x = 0.9$, the hydrogen content stabilizes at 1.2 H/f.u. For pure vanadium, the hydrogen uptake increases up to 1.6 H/f.u. after 6 min of milling and then it gradually decreases down to 1.2 H/f.u. over 30 min. It is worth noting that V-rich mixtures (*e.g.* $x = 0.8$ and 1) start reacting with hydrogen earlier than Ti-rich ones (*e.g.* $x = 0$ and 0.25).

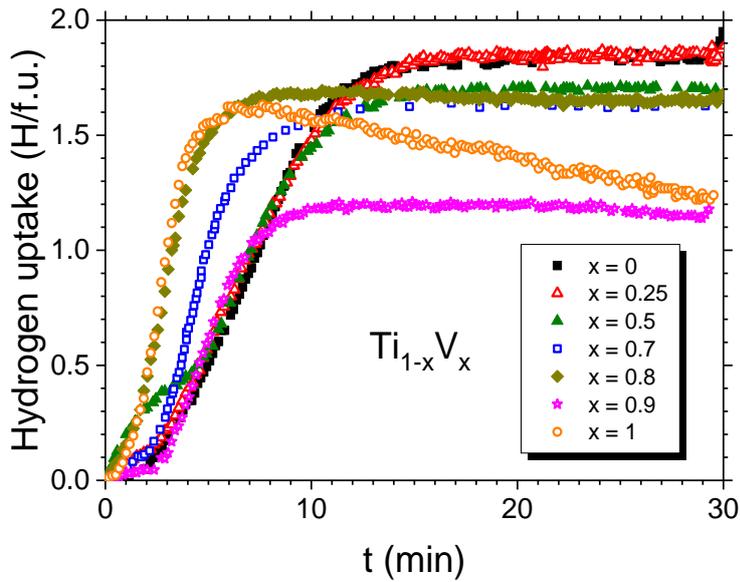

*Figure 1: Hydrogen uptake curves of $Ti_{1-x}V_x$ powder mixtures during ball milling under hydrogen gas*

The atomic ratio between Ti and V of the as-milled powders was analyzed by ICP-OES. Results are given in Table S1 as Supplementary Information (SI). The measured Ti:V metal ratio matches well with the nominal composition of the reactants.

Figure 2 : XRD patterns of $Ti_{1-x}V_x$ hydrides synthetized by mechanochemistry under hydrogen gas. Diffraction peaks from fcc ($Ti_{1-x}V_x$) dihydride phases are marked with solid triangles. Vertical dashed lines stand for fcc-$TiH_2$ (x = 0) as a guide to the eye for noting fcc peak shifts. Diffraction peaks from bct VH (x = 1) are marked with empty triangles.Figure 2 displays the XRD diffraction patterns of the synthetized $Ti_{1-x}V_x$ hydrides. The full set of patterns can be indexed with three phases: face-centered cubic (fcc, S.G. $Fm\overline{3}m$), body-centered tetragonal (bct, S.G. $I4/mmm$) and body-centered cubic Fe (bcc, S.G. $Im\overline{3}m$). The occurrence of the latter phase, characterized by a main peak at $2\theta_{Cu} = 44.5°$, is attributed to vial wearing by the high hardness of the hydride powder [26,27], whereas fcc and bct phases are assigned to the formation of $Ti_{1-x}V_x$ di- and mono-hydrides, respectively.

For samples with $x \leq 0.8$ main diffraction peaks relate to the fcc lattice of $TiH_2$ [28] and $VH_2$[29] dihydrides which share an isostructural fluorite-type structure. Peak positions shift towards higher angles, *i.e.* cell volume shrinks on increasing V content. In contrast, for $x \geq 0.9$ the bct phase becomes predominant and the single-one detected for $x = 1$. It is attributed to the formation of $\beta_2$-VH monohydride. The occurrence of this tetragonal phase has been reported by Orimo *et al.* [30] on nanostructured vanadium-hydrogen samples synthetized by same method as here, but at lower hydrogen pressure (< 1 MPa).

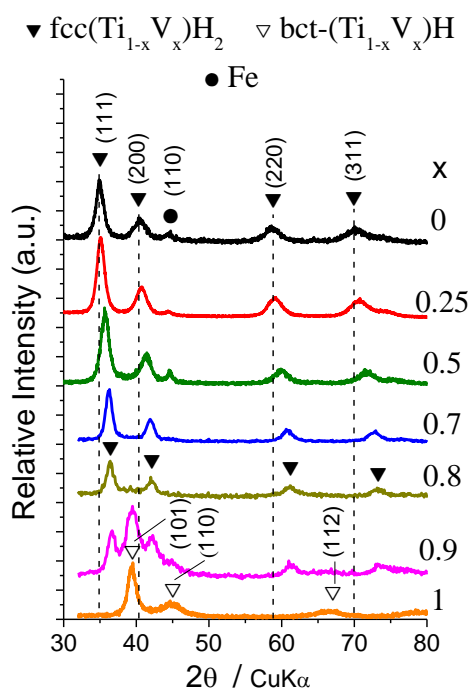

*Figure 2 : XRD patterns of $Ti_{1-x}V_x$ hydrides synthetized by mechanochemistry under hydrogen gas. Diffraction peaks from fcc ($Ti_{1-x}V_x$) dihydride phases are marked with solid triangles. Vertical dashed lines stand for fcc-$TiH_2$ (x = 0) as a guide to the eye for noting fcc peak shifts. Diffraction peaks from bct VH (x = 1) are marked with empty triangles.*

To get further information on phase content and microstructural properties of the synthetized samples, Rietveld analyses were performed for all patterns. A typical graphical output of the refinement obtained for $x = 0.7$ is given in Figure S1 showing a satisfactory fit between experimental and calculated patterns. Crystallographic data for all samples are gathered in Table 1.

Fe contamination below 7 wt% was detected by ICP for $Ti_{1-x}V_x$ samples with $x \leq 0.5$. For V-rich samples, Fe peaks could not be detected. This may result from either lower abrasion property of the synthetized hydrides or Fe solubility in the hydride phases. Fcc $Ti_{1-x}V_xH_2$ is the major and unique hydride phase for $x = 0.7$ and 0.8. Its crystallite size, $L$, ranges from 5 to 12 nm with no clear dependence on metal ratio. In contrast, the microstrain steadily decreases from 1.6 to 0.4 % when vanadium content increases from $x = 0$ to 0.8. The bct monohydride $Ti_{1-x}V_xH$ appears at $x = 0.9$ and is the unique phase for $x = 1$. It has high microstrain with crystallite size in the nanometric range, $L = 6$ nm.

Table 1 Crystallographic data of $Ti_{1-x}V_x$ hydrides synthetized by mechanochemistry under hydrogen gas. Cell parameters micro-strain, crystallite size (L) and Rietveld agreement factors ($R_B$, $R_{wp}$) are given. Standard deviations refereed to the last digit are given in parenthesis.

| Sample (x) | Phase | S.G. | Cell parameters a (Å) | b (Å) | c (Å) | μ-strain (%) | L (nm) | Content (wt.%) | $R_B$ (%) | $R_{wp}$ (%) | $\chi^2$ |
|---|---|---|---|---|---|---|---|---|---|---|---|
| 0 | fcc | Fm-3m | 4.451(2) | 4.451(2) | 4.451(2) | 1.6(1) | 8(1) | 93(2) | 3 | 13 | 1.2 |
|   | Fe | Im-3m | 2.883(1) | 2.883(1) | 2.883(1) | 1* | 5(1) | 7(2) | 4 |    |     |
| 0.25 | fcc | Fm-3m | 4.416(1) | 4.416(1) | 4.416(1) | 1.6(1) | 12(2) | 97(1) | 4 | 8 | 2.2 |
|   | Fe | Im-3m | 2.881(1) | 2.881(1) | 2.881(1) | 1* | 12* | 3(1) | 10 |   |   |
| 0.5 | fcc | Fm-3m | 4.373(2) | 4.373(2) | 4.373(2) | 1.2(1) | 6(1) | 96(2) | 6 | 18 | 1.6 |
|   | Fe | Im-3m | 2.881(1) | 2.881(1) | 2.881(1) | 0.4* | 8* | 4(2) | 9 |   |   |
| 0.7 | fcc | Fm-3m | 4.324(1) | 4.324(1) | 4.324(1) | 0.9(1) | 9(1) | 100 | 6 | 14 | 1.6 |
| 0.8 | fcc | Fm-3m | 4.296(2) | 4.296(2) | 4.296(2) | 0.4(2) | 5(1) | 100 | 13 | 15 | 1.5 |
| 0.9 | fcc | Fm-3m | 4.297(3) | 4.297(3) | 4.297(3) | 0.4* | 5* | 25(2) | 11 | 16 | 2.6 |
|   | bct | I4/mmm | 2.942(3) | 2.942(3) | 3.679(5) | 3.1(1) | 6* | 75(2) | 10 |   |   |
| 1 | bct | I4/mmm | 2.900(2) | 2.900(2) | 3.879(3) | 1.4(1) | 6(1) | 100 | 5 | 22 | 2.2 |

*Values were fixed to ensure fit stability mostly due to low phase amount.

### 3.2 Electrochemical properties of Ti-V hydrides in Li-ion half-cells

Figure 3 shows the potential profile during the first galvanostatic cycle of representative hydrides $Ti_{1-x}V_x$ ($x = 0, 0.25, 0.7$ and $0.8$). These compounds were selected as they are H-rich dihydrides and, regardless Fe contamination, single-phase hydrides (Figure 1, Table 1). All potential profiles exhibit a shoulder on discharge (*i.e.* on lithiation) at $E \sim 0.9$ V vs. Li$^+$/Li that, as later discussed, is attributed to the formation of a solid electrolyte interface (SEI). For potentials below 0.5 V vs. Li$^+$/Li profile trends are dissimilar for Ti-rich ($x = 0$ and 0.25) and V-rich ($x = 0.7$ and 0.8) compositions. For Ti-rich hydrides, the potential gradually decreases down to ~ 0.25 V followed by a sloping plateau. The average plateau potential increases with V-content from 0.08 to 0.18 V vs. Li$^+$/Li for $x = 0$ and 0.25, respectively. For V-rich compositions, a local minimum juncture in the potential profile is first observed at ~ 0.43 V vs. $Li^+/Li$ that is followed by a flat plateau at ~ 0.45 V vs. $Li^+/Li$ which gradually decreases down to ~ 0.2 V vs. $Li^+/Li$ and then more abruptly to the cut-off potential of 0.02 V vs. $Li^+/Li$. Independently of the profile shapes, for all hydrides the signals below 0.5 V vs. $Li^+/Li$ can be tentatively attributed to conversion reactions between hydrides and lithium. Indeed, by considering that the SEI reaction extends in all cases to 0.7 Li, one remarks that, below 0.5 V vs. $Li^+/Li$, roughly 2 additional atoms of Li (*i.e.* $\Delta y$ ~ 2) react with every hydride. The conversion reaction, in which one Li atom reacts by each H atom, can be expressed by:

$$Ti_{1-x}V_xH_y + yLi^+ + ye^- \rightarrow Ti_{1-x}V_x + yLiH \qquad (R1)$$

Full lithiation is typically achieved with 2.7 Li, which corresponds to a discharge capacity $Q_{discharge}$ of ~ 1500 mAh $g^{-1}$. For all electrodes, the first delithiation is limited to $\Delta y$ ~ 0.6 (*i.e.* $Q_{charge}$ ~ 300 mAh $g^{-1}$) and no plateau potential is observed. The reversible capacity gradually decreases on further cycling as can be observed in Figure S2.

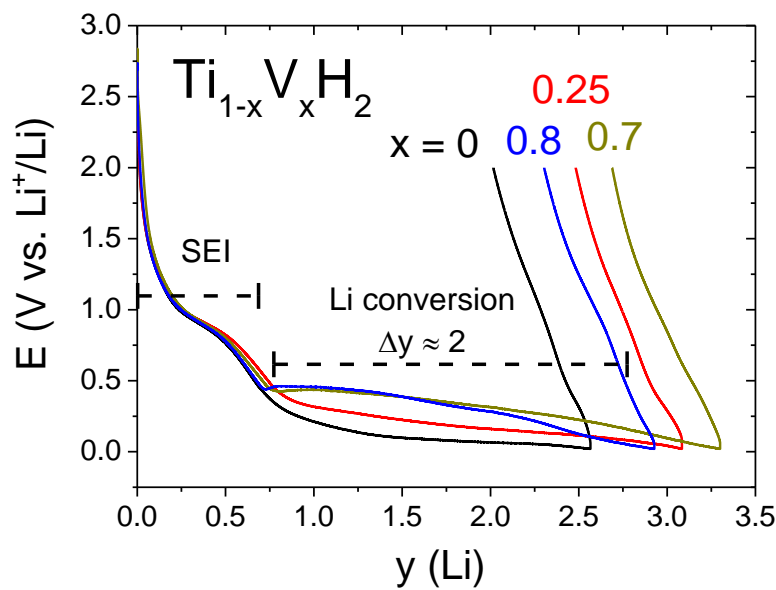

*Figure 3 : Potential profiles of selected dihydrides $Ti_{1-x}V_xH_2$ during the first galvanostatic cycle in Li-ion half-cell.*

Further analysis of electrochemical properties was gained by cyclic voltammetry. As a representative example, Figure 4a shows the cyclic voltammogram of $Ti_{0.75}V_{0.25}H_2$ electrode during the first lithiation/delithition cycle. Two cathodic peaks at 0.9 and 0.05 V vs. Li$^+$/Li are observed upon lithiation. They are assigned to SEI and conversion reaction (R1), respectively. No clear anodic peaks are observed on delithiation which confirm the limited reaction reversibility. The peak potential and current assigned to the Li–conversion reaction clearly depends on the hydride composition as shown in Figure 4b. On increasing V-content, the peak potential gradually increases. This concurs with the destabilization effect of vanadium already observed during galvanostatic cycling. In addition, the peak current decreases with V-content, suggesting that the kinetics of the conversion reaction slow down on increasing V-content.

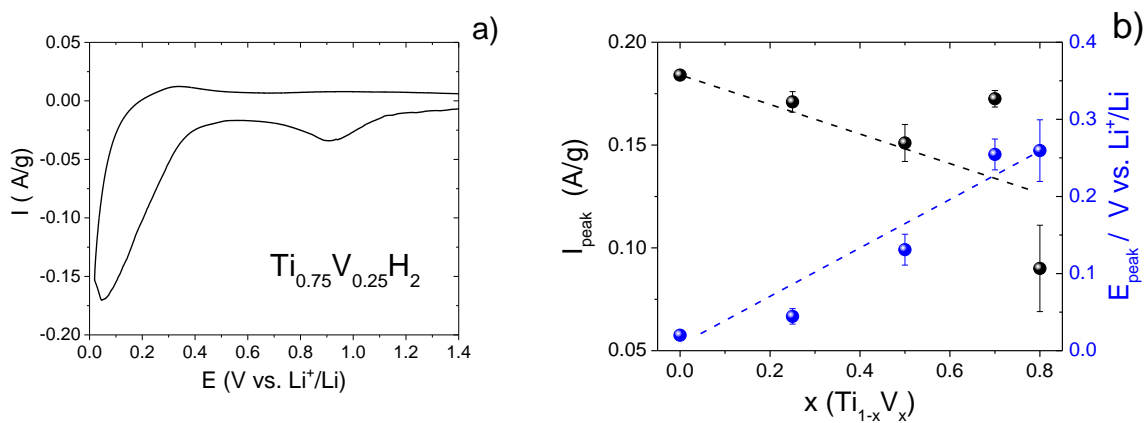

*Figure 4 : Cyclic voltammetry of dihydrides $Ti_{1-x}V_x H_2$ for the first reduction/oxidation cycle. a) voltammogram for x = 0.25. b) cathodic peak potential(blue) and intensity (black) for all hydride compounds (Dashed lines are guides to the eye).*

## 4. Discussion
### 4.1 Formation and stability of Ti-V hydrides

Hydrogen uptake curves of $Ti_{1-x}V_x$ powders during mechanochemical synthesis (Figure 1) and structural analysis of the resulting hydrides (Table 1) show that dihydride compounds are formed for V-contents below $x = 0.8$, while monohydrides occur above this limit. These results concur with previous findings of Dunlap *et al* [31] and Orimo *et al* [30] for the mechanosynthesis of binary hydrides of the pure Ti and V metals for initial pressure of 0.1 and 1 MPa, respectively. Formation of dihydride $TiH_2$ is expected since under atmospheric pressure this compound is thermodynamically stable below 300 °C [32]. In contrast, formation dihydride $VH_2$ by mechanochemistry is more intricated. The desorption equilibrium pressure of fcc-$VH_2$ is ~ 0.2 MPa at room temperature [29] and therefore $VH_2$ is unstable at atmospheric pressure (*i.e.* at the conditions used by Dunlap [31]). However, at the conditions used in this study, with a working pressure of ~ 8 MPa, VH$_2$ should be stable unless the reaction temperature greatly excess RT. Indeed, a close inspection of the hydrogen uptake curve for V ($x = 1$) in Figure 1 reveals a two-step process with hydrogen absorption up to 1.6 H/f.u. during the first 6 min of milling followed by a decrease on hydrogen content down to 1.2 H/f.u. on prolonged milling. The absorption step implies a significant formation of VH$_2$, while the desorption one requires VH$_2$ decomposition. This hydrogen uptake/release behaviour of vanadium during milling can be understood by considering the heating of the powder due to collision impacts between milling balls and powder material. To this respect, the recorded vial temperature increases gradually during milling (Figure S3). This temperature is known to be much lower than the local temperature at the collision impact. Indeed, according to the thermodynamics of the V-H system [29], the desorption equilibrium pressure is 8 MPa at 115 °C for VH$_2$. By comparison with previous mechanochemical studies [33], this temperature can be reasonably attained at the milling conditions used in this work.

Let's consider now the ternary Ti-V-H system. As evidenced by XRD data (Figure 1), formation of single-phase Ti$_{1-x}$V$_x$H$_2$ compounds occur for V-contents below $x = 0.8$. Their structural properties and thermodynamic stability are worth to discuss. The lattice constant of the synthetized fcc-Ti$_{1-x}$V$_x$H$_2$ compounds as obtained from Rietveld refinements is displayed in Figure 5. It decreases linearly with V-content, in close agreement with Vegard's law and following the same trend as dihydrides Ti$_{1-x}$V$_x$H$_2$ formed by classical solid-gas reaction after the synthesis of Ti-V alloys by melting [34]. It evidences that these dihydrides are pseudo-binary Ti$_{1-x}$V$_x$H$_2$ compounds, with Ti and V atoms sharing the same 4*a*-Wyckoff site, *i.e.* sitting at the corners and center of each cube face of the fluorite-type structure. Thus, (Ti,V) behaves as a pseudobinary atom, which leads the lattice contraction with increasing V content of both the dihydride and the parent intermetallic compounds [35,36]. This lattice contraction derives from the smaller atomic radius of V as compared to Ti (r$_V$ = 1.316 Å, r$_{Ti}$ = 1.461 Å) [37].

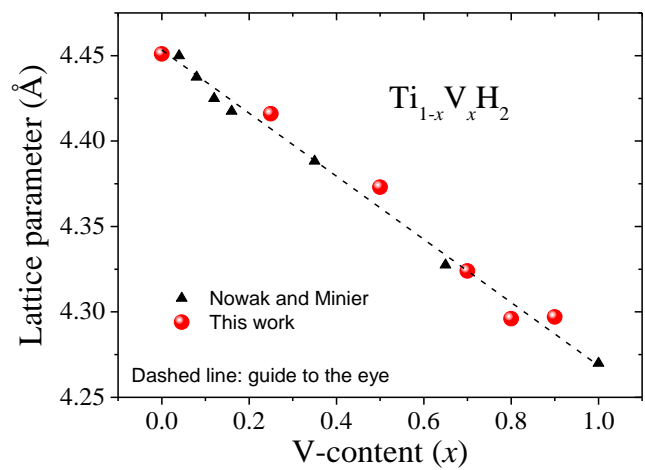

*Figure 5 : Lattice parameter of the cubic fcc-structure of $Ti_{1-x}V_xH_2$ hydrides. Data obtained in this work by mechanochemistry are compared to samples prepared by melting and hydrogenated by classical solid-gas reaction* [34].

The thermodynamic stability of these $Ti_{1-x}V_xH_2$ compounds can be now discussed at the light of geometric and electronic models. As empirically shown by Lundin *et al.* [38] and Diaz *et al.* [39] for many hydride forming compounds, the hydride stability decreases with shrinking of the intermetallic cell volume. Then, according to the previous structural analysis (Figure *5*), the stability of dihydrides $Ti_{1-x}V_xH_2$ are expected to decrease with V content. Indeed, Kagawa *etal.* [40] and Massicot *et al.* [19] have shown that the enthalpy of formation of these dihydrides linearly increases with V-content leading to hydride destabilization. This trend can be also examined at the light of Miedema's rule that states that the more stable the intermetallic compound, the less stable is its hydride. As determined from first-principles calculations [41], the enthalpy of formation of bcc Ti-V alloys decreases with V-content and, according to Miedema's rule, corroborating the destabilization of the hydride.

### 4.2 Conversion reaction of dihydrides $Ti_{1-x}V_xH_2$ with lithium

The conversion properties of pseudobinary Ti-V dihydrides are here reported for the first time but previous studies exist for the binary compounds $TiH_2$ and $VH_2$. $TiH_2$ has been cycled in Li half-cells at RT [7] and at 125°C [42] in combination with liquid and solid electrolytes, respectively. In both cases, the lithiation profile of $TiH_2$ is characterized by a gradual potential decrease between 0.5 and 0.15 V vs $Li^+/Li$ with the reaction of 1 Li atom, leading to the transformation of fcc-$TiH_2$ to face-centered-orthorhombic (fco) monohydride -TiH, followed by a plateau at 0.15 V due to the reaction of TiH with an additional Li atom to form hcp-Ti. As for $VH_2$, Ichikawa's group has reported the conversion reaction with Li, using a working electrode of a molar ratio 1:3 V:LiH due to the instability of $VH_2$ at normal conditions of $P$ and $T$ [43]. A first plateau at 0.65 V vs. $Li^+/Li$ was observed upon delithiation and attributed to $VH_2$ formation though it could not be confirmed by XRD due to the instability of $VH_2$. On lithiation, a plateau between 0.6 and 0.45 V vs. $Li^+/Li$ was detected followed by a slopping branch leading to the formation of pure vanadium.

The potential profile previously described for $TiH_2$ conversion matches well with the results reported in section 3.2 (Figure 3) with the reaction of $y = 1.9$ and 2.3 atoms of Li in the 0 – 0.5 V range for Ti-rich samples $x = 0$ and 0.25, respectively. The reaction extent agrees well with the conversion reaction of dihydrides (y ~ 2), considering the accuracy of measurements. It should be noted that the plateau potential is higher for $x = 0.25$ than for $x = 0$ due to vanadium-induced destabilization. In contrast, for V-rich samples ($x = 0.7$ and 0.8) a flat plateau is observed at high voltage (~ 0.45 V) followed by a gradual potential decrease in close agreement with Ichikawa's results for $VH_2$ [43]. Again, the plateau potential is observed to increase with V-content, with the reaction of $y = 2.6$ and 2.2 Li in the 0 – 0.5 V range for samples $x = 0.7$ and 0.8, respectively. Such excess of Li reaction as compared to the full conversion of for dihydride compounds, $y = 2$, might be explained by minor contribution of carbon lithiation (see Figure S5) or SEI formation due to the generation of fresh surfaces during the conversion reaction.

The potential at which the conversion reaction between $Ti_{1-x}V_xH_2$ compounds and lithium should occur can be calculated using the Nernst equation:

$$E_{Nernst} \ (vs.Li^+/Li) = \frac{1}{F}\left[\frac{\Delta_f G^o(Ti_{1-x}V_xH_2)}{y} - \Delta_f G^o(LiH)\right] \quad \text{(Eq. 1)}$$

where $F$ stands for the Faraday constant (96485 C $mol^{-1}$), $y$ is the number of Li ions that participate in the overall conversion reaction ($y = 2$ for dihydrides in R1), and $\Delta_f G°$ is the standard Gibbs free energy of formation of reactants and products. This energy is tabulated for the binary compounds: -69.96, -91.67 and 3.52 kJ $mol^{-1}$ for LiH, $TiH_2$ and $VH_2$, respectively [44]. Based on the geometric and electronic considerations discussed in the previous section, the Gibbs free energy of formation of pseudobinary $Ti_{1-x}V_xH_2$ hydrides can be reasonably approximated by a linear equation comprising the binary end-compounds $TiH_2$ and $VH_2$ :

$$\Delta_f G^o(Ti_{1-x}V_xH_2) = -91.67 + 95.19 \ x \quad \text{(Eq.2)}$$

Thus, the conversion potentials can be calculated combining Eqs. 1 and 2 and compared with the experimental ones. The latter were taken from Figure 3 and arbitrarily selected at $y = 1$ as a common reference for the starting of the conversion reaction for all electrodes, occurring after SEI formation. The comparison between calculated and experimental data is displayed in Figure 6. Both potentials follow the same trend as a function of V-content. Contrary to calculated data, experimental ones are far from equilibrium and thus shifted towards lower values due to the imposed current during galvanostatic cycling. Indeed, the difference between them should represent the discharge overpotential which increases with vanadium content, in agreement with cyclic voltammetry data (Figure 4b) that indicate that the reaction kinetics slow down with V-content.

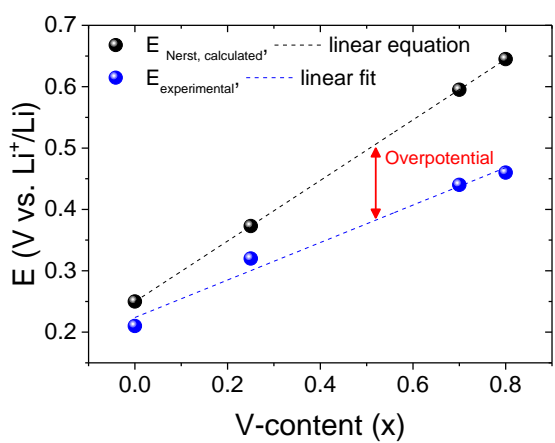

*Figure 6 : Calculated and experimental potentials of the conversion reaction between $Ti_{1-x}V_xH_2$ and Li. The difference between both potentials, labelled as overpotential, is expressly shown.*

To get further insight on reaction mechanisms, differential capacity plots d$Q$/d$V$ of the potential profiles (Figure 3) have been determined and results are displayed in Figure 7 for the potential range $0 \leq E \leq 0.6$ (V vs. $Li^+$/Li) and in Figure S4 when $E$ is extended up to 1.2 V vs. $Li^+$/Li. For all electrodes, a common peak at 0.9 V vs. $Li^+$/Li is observed. This signal is not observed on blank experiments using electrodes only formed by the CMC binder and carbon conducting additive (Figure S5). Therefore, it is assigned to an irreversible secondary reaction with participation of the metal hydride but independent of the alloy composition. This signal is attributed to the formation of a solid electrolyte interface over the metallic hydride. The SEI formation potential is in agreement with previous reports [17,45], though it was assigned to SEI formation over conductive carbon, a fact that is not supported by the present observations.

For potentials below 0.5 V vs $Li^+/Li$, differential plots clearly depend on the hydride composition. For $TiH_2$ sample ($x = 0$), the derivative curve departs from the horizontal line at ~ 0.4 V vs. $Li^+/Li$, with a gradual increase on intensity showing a minor peak at 0.09 V vs. $Li^+/Li$ and a major one at 0.06 V vs. $Li^+/Li$. This signature concurs with the three-step conversion reaction reported by Oumellal *et al* [17], starting with the reaction between the solid solution of hydrogen in the δ-fcc phase (δ-$TiH_{2-\varepsilon}$; $0 \leq \varepsilon \leq 0.34$) and lithium, followed by a two-phase reaction involving δ-fcc $TiH_{1.64}$ and orthorhombic δ-fco-TiH, and ending with the conversion reaction from the latter monohydride to hcp-Ti metal. For $x = 0.25$, though the starting of the conversion reaction occurs as well at ~ 0.4 V vs. $Li^+/Li$, pointing to the solid solution mechanism as for $x = 0$, the d$Q$/d$V$ plot shows clear differences with the shift of the low-potentials peaks to higher values (0.13 and 0.16 V vs. $Li^+/Li$) and the appearance of a novel peak at 0.29 V vs. $Li^+/Li$. Peak shift to higher potentials can be assigned to the destabilization of δ-titanium hydride due to partial vanadium substitution. In contrast, the novel peak at 0.29 V vs. $Li^+/Li$ indicates a change in the reaction path which can be understood at the light of d$Q$/d$V$ plots for V-rich samples. Indeed, for $x = 0.7$ and 0.8, the start of the conversion reaction occurs abruptly at 0.47 V vs. $Li^+/Li$ with a discontinuity in the differential plots assigned to a nucleation and growth mechanism [46,47]. The low solubility of hydrogen in δ–fcc vanadium dihydride [29] accounts for this, with the nucleation and growth of $β_2$-bct vanadium monohydride according to the conversion reaction $VH_2 + Li \rightarrow VH + LiH$ in agreement with recent results of Matsumura *et al* [43]. After the voltage peak of nucleation, the growth of vanadium monohydride is evidenced by a d$Q$/d$V$ peak located at slightly higher potential with increasing V-content: 0.42 and 0.44 V vs. $Li^+/Li$ for $x = 0.7$ and 0.8 V vs. $Li^+/Li$, respectively, showing again the destabilization effect of this substitution. Finally, at lower potentials, the differential d$Q$/d$V$ peak located at ~ 0.29 V vs. $Li^+/Li$ can be assigned to conversion reactions from bct-VH to a hydride with lower hydrogen content. Indeed, at room temperature and atmospheric pressure, the V-H phase diagram exhibits two different β-phases, namely $β_1$ and $β_2$, with slightly different structures and hydrogen content: monoclinic $VH_{0.5}$ and tetragonal $VH_{0.8}$ for $β_1$ and $β_2$ respectively [48].

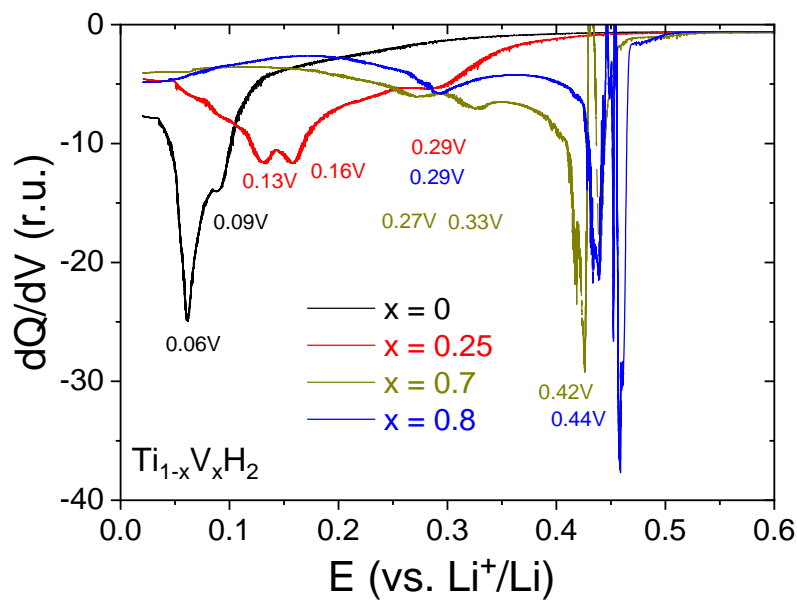

*Figure 7 : Differential capacity plots as a function of voltage for selected dihydrides $Ti_{1-x}V_x H_2$ during the first galvanostatic cycle in Li-ion half-cell. Peak potentials are expressly given.*

The peculiar behaviour of the conversion reaction of V-rich hydrides with lithium was further investigated by *operando* XRD in a specially designed cell for in-situ experiments [25,49]. For this purpose, $Ti_{0.3}V_{0.7}H_2$ hydride was selected. The potential lithiation profile and collected XRD data are shown in detail in Figure S6. No change in XRD patterns is observed for $y < 0.8$ Li and potentials above 0.5 V vs. Li$^+$/Li, confirming that side reactions, likely SEI formation, occur within this region. Next, at lower potentials, the intensity of fcc-$Ti_{0.3}V_{0.7}H_2$ diffraction peaks gradually decreases, while broad peaks attributed to formation of bct-$Ti_{0.3}V_{0.7}H_z$ hydride with variable hydrogen content, $z$, appear. Diffraction patterns during this conversion reaction were sequentially refined by the Rietveld method and results are given in Figure 8. The amount of bct-$Ti_{0.3}V_{0.7}H_z$ hydride continuously increases with lithiation progress at the expense of the fcc-$Ti_{0.3}V_{0.7}H_2$ hydride (Figure 8a). During this phase transformation, the lattice parameter of the dihydride fcc phase remains almost constant ($a_{fcc} = 4.330 \pm 0.007$ Å) while those of bct-$Ti_{0.3}V_{0.7}H_z$ gradually decrease during lithiation pointing to a progressive diminution in hydrogen content. The constant value of $a_{fcc}$ during the conversion reaction agrees with the low solubility of hydrogen in fcc-$VH_2$ and the nucleation and growth of bct-VH in fcc-$VH_2$ previously discussed (Figure 7). For the latter phase, the $a_{bct}$ parameter gradually decreases from 3.138 to 3.067 Å (for $y = 0.86$ and 1.95, respectively), whereas the $c_{bct}$ parameter undergoes an abrupt drop at $y \sim 1.25$ from 3.738 to 3.430 Å. This drop concurs, as previously discussed, with the reported formation of two different $\beta_1$ and $\beta_2$ hydride phases with dissimilar hydrogen content in the V-H system [48]. Their crystal structure might differ, but it could not be here resolved due to their poor crystallization. Moreover, it should be noted that the tetragonal distortion of the bct-phase reduces on lithiation, being only $(c/a)_{bct} = 1.07$ at $y = 1.95$. Indeed, the lattice parameters at the end of the lithiation process approach to those of the cubic bcc alloy $Ti_{0.3}V_{0.7}$, with $a_{bcc} = 3.106$ Å [35], suggesting that the conversion reaction gradually comes to its end following a solid-solution process with the formation of the parent bcc alloy $Ti_{0.3}V_{0.7}$ and LiH. The occurrence of LiH could not be confirmed due to the low electronic density of its constituent elements.

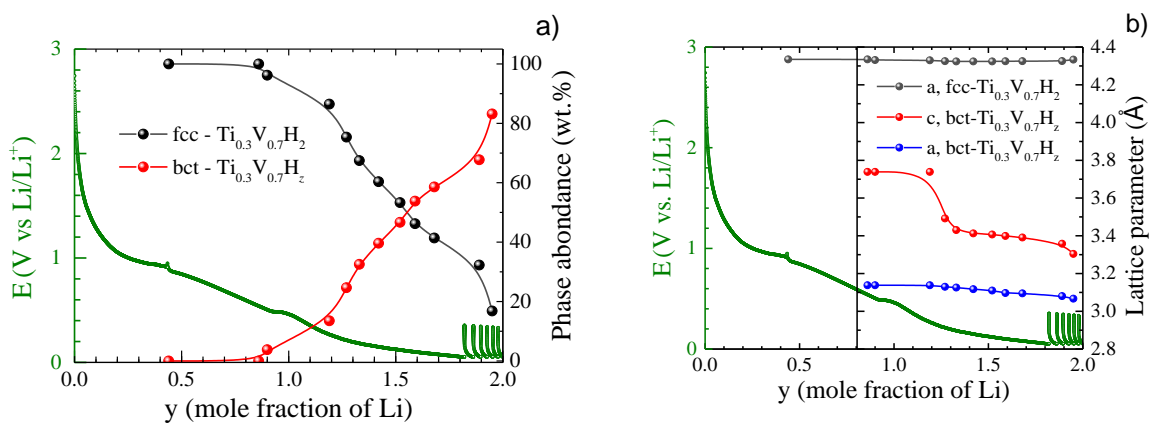

*Figure 8 : Potential profile and sequential crystallographic data of $Ti_{0.3}V_{0.7}H_2$ hydride during operando lithiation. a) Evolution of phase content as a function of lithiation, b) Evolution of lattice parameters for the two detected fcc and bct hydride phases.*

To end up with the conversion reaction with lithium, it is worth to briefly discussing on the limited reversibility for all studied compounds. This property concurs with previous findings of Aymard *et al*, which reported that the conversion reaction of $TiH_2$ at RT is not reversible after the first lithiation [17]. However, reversibility of $TiH_2$ has been later attained by different strategies, such as increasing temperature to 125°C, using Ti-LiH composites as starting materials[50], or embedding $TiH_2$ in a nanostructured Mg/LiH matrix [45]. Though the exact reason of this reversibility issue remains unsolved, both kinetic and interfacial concerns are suspected to be at the origin of this phenomenon.

### 4.3 Conversion reaction of dihydrides $Ti_{1-x}V_xH_2$ with sodium

The fact that $Ti_{1-x}V_xH_2$ compounds react with Li at a potential which can be tuned at wish as a function of their composition constitutes a major result of this research. This paves the way to tailor the electrode potential to adapt its operation to be compatible with specific electrolytes or alternative chemistries. For instance, as regards the latter, the conversion reaction between $TiH_2$ and Na is not expected to occur since its reaction potential ($E_{Nernst}$ = 0.25 V vs. Li$^+$/Li) is negative ($E_{Nernst}$ = -0.12 V vs. Na$^+$/Na). Therefore, Na plating should occur before its conversion in $TiH_2$|Na half-cell. Thus, based in the previous study on Li-ion half-cell, the V-rich compound $Ti_{0.3}V_{0.7}H_2$ with a high plateau potential of $E_{Nernst}$ = 0.22 V vs. Na$^+$/Na was selected to investigate the conversion mechanism during sodiation.

$MgH_2$ is the only metal hydride studied so far as electrode material in Na-ion batteries. It has been reported to achieve a capacity of sodiation of 350 mAh g$^{-1}$ (17 % of the theoretical capacity) at 0.15 V vs. Na$^+$/Na with very limited reversibility [20]. The mechanism of this limited sodiation reaction was not clearly established. For the selected $Ti_{0.3}V_{0.7}H_2$ compound, the potential profiles of galvanostatic sodiation and lithiation are superimposed in Figure 9. For easier comparison, potential values are converted *vs*. normal hydrogen electrode (NHE). A plateau at -2.69 V vs. NHE (almost 0 V vs. Na$^+$/Na) is observed in the case of Na-ion half-cell. This value is lower than in Li-ion configuration (-2.59 V vs. NHE) meaning higher polarization for sodium that for lithium conversion. This difference is tentatively assigned to slower reaction kinetics for sodiation as compared to lithiation. It is worth to remark that in both cases a local minimum potential is observed which is attributed to similar nucleation and growth mechanisms. A capacity of 300 mAh g$^{-1}$ has been obtained for sodiation which corresponds to 28 % of the theoretical capacity, much higher than that obtained with $MgH_2$ [20]. To gain stronger evidence that the conversion reaction takes place, a forced sodiation with GITT method (Figure S7) has been realized in half-cell. Sodiation was imposed for 4 h for the first 4 steps, while next ones were limited either to 2 h or by the cut-off potential. After completing a total reaction of $y$ ~ 1.5 Na, an ex-situ XRD analysis has been performed on the electrode after rinsing with EC and drying in glove box. The diffraction peaks of NaH are detected. Evaluation of NaH mass fraction by Rietveld analysis (Figure S8) leads to a reaction of 0.95 Na. This amount agrees with the potential profile (Figure S7) when considering that the conversion reaction is preceded by SEI formation up to $y$ ~ 0.6 Na and ends at $y$ ~ 1.5 Na. Complementary ICP analyses (see SI for details) of the sodiated electrode composite showed a higher content of sodium, equivalent to 1.25 mol NaH for one mol of $Ti_{0.3}V_{0.7}H_2$. The difference between ICP and XRD analysis can be explained by SEI formation. In conclusion, the galvanostatic potential profile, XRD and ICP demonstrate the formation of NaH in the sodiated electrode according to the follow reaction:

$$\text{Ti}_{0.3}\text{V}_{0.7}\text{H}_2 + 0.9\ \text{Na}^+ + 0.9\ \text{e}^- \rightarrow 0.9\ \text{NaH} + 0.55\ \text{Ti}_{0.3}\text{V}_{0.7}\text{H}_2 + 0.45\ \text{Ti}_{0.3}\text{V}_{0.7} \quad (R2)$$

Therefore, the conversion reaction of Ti-V hydrides in a Na-ion battery is unambiguously demonstrated here for a half-cell configuration. After a low reversibility observed during the first cycle (0.15 Na during desodiation against 0.55 in sodiation plateau in Figure 9), further cycles involve very limited capacity. Beside the problem of interfacial contact, the kinetics should be the main factor considering the diffusion of the heavy Na$^+$ comparing to Li$^+$.

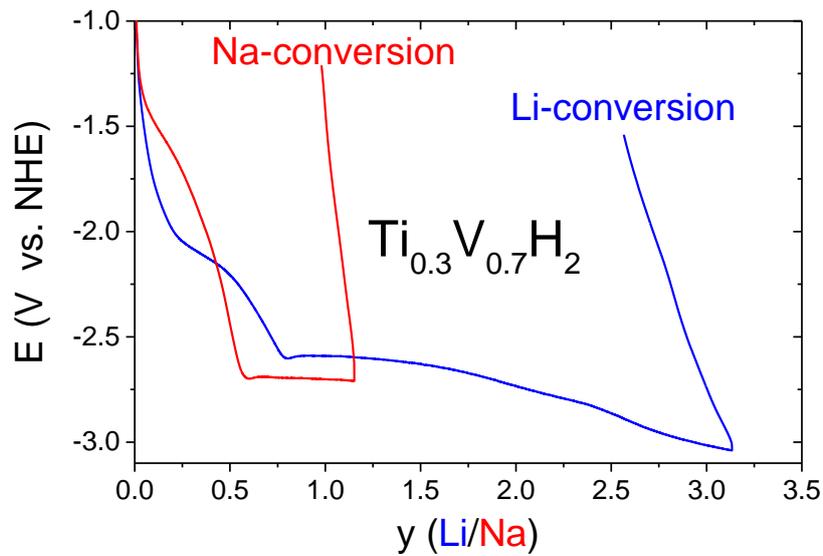

*Figure 9 : Comparison of the potential profiles of dihydride Ti$_{0.3}$V$_{0.7}$H$_2$ during the first galvanostatic cycle in Li-ion and Na-ion half cells*

## 5. Conclusion

We investigated the formation of Ti$_{1-x}$V$_x$H$_2$ hydrides (0 ≤ $x$ ≤ 1) and the feasibility of their conversion reaction with Li and Na conversion in half cell. For Ti-rich samples with $x \leq 0.8$ the formation of fcc dihydride phase was successfully achieved by mechanochemistry with a linear decrease of lattice parameter with V content. In contrast, for $x \geq 0.9$ the bct monohydride phase becomes predominant and the single-one detected for $x = 1$. The full Li-conversion in the 0 – 0.5 V vs Li$^+$/Li range was obtained, and the destabilization effect of vanadium was confirmed by the increase of the potential with V content during Li-conversion, in line with calculated Nernst potential from thermodynamical features. Moreover, reaction paths for Li-conversion depend on the hydride composition. For Ti-rich compounds, the conversion reaction starts with a solid solution behaviour followed by a potential plateau attributed to the hydrogen depletion from titanium monohydride to titanium. In contrast, for V-rich compounds, the conversion reaction starts with a potential plateau assigned to the transformation of vanadium dihydride to monohydride followed by a decreasing potential. This decrease is attributed to the formation of several hydride phases with low hydrogen content leading at the end to the formation of vanadium. The potential at the beginning of the plateau, lower than the theoretical value calculated at equilibrium, indicates an important overpotential certainly coming from kinetic limitation. Indeed, a weak and fading reversibility was observed during galvanostatic cycling. The Na-conversion with NaH formation upon sodiation was established for $Ti_{0.3}V_{0.7}H_2$ but with detrimental kinetics limitations that prevent a reversible conversion at RT. To be able to use hydride as efficient anodes for *M*-ion battery, deeper investigation will be necessary to decipher and overcome the weak reversibility of the Li and Na conversion close to room temperature.

### Acknowledgements

The authors thank the Bachelor student Xavier Barhdadi for his experimental work and the French-Australian network IRN-FACES for financial support.

This manuscript honours the memory our colleague Michel Latroche who significantly contributed to this research and suddenly passed away.

### CRediT authorship contribution statement

**Fermin Cuevas**: Conceptualization, Formal analysis, Writing - original draft, Writing - review & editing, Funding acquisition, Supervision. **Barbara Laïk**: Data curation; Formal analysis; Funding acquisition; Investigation; Methodology; Supervision; Writing - review & editing. **Junxian Zhang**: Data curation; Formal analysis; Funding acquisition; Investigation; Methodology; Supervision; Writing - review & editing. **Mickaël Mateos**: Formal analysis, Writing – review & editing. **Jean-Pierre Pereira-Ramos**: Supervision, Writing – review & editing. **Michel Latroche**: Conceptualization, Supervision.

# Supplementary Information

# Mechanochemical synthesis of pseudobinary Ti-V hydrides and their conversion reaction with Li and Na

Fermin Cuevas[*], Barbara Laïk, Junxian Zhang, Mickaël Mateos, Jean-Pierre Pereira-Ramos, Michel Latroche[†]

Univ Paris-Est Creteil, CNRS, ICMPE (UMR 7182), 2 rue Henri Dunant, F-94320 Thiais, France

The chemical composition of the as-milled $Ti_{1-x}V_xH_y$ samples was determined by ICP-OES, as concerns metal content, and by the hydrogen pressure drop during ball-milling synthesis. Results are gathered in Table S1.

*Table S1. Chemical composition of as-milled $Ti_{1-x}V_xH_y$ samples. H-content, y, measured with Evicomagnetics device. Metal content determined by ICP-OES analysis (Fe-contamination not considered).*

| Sample | Metal-content | | H-content |
|---|---|---|---|
| $x$ | at.%Ti | at.% V | $y$ |
| 0 | - | - | 1.8(2) |
| 0.25 | 72.6 | 27.3 | 1.8(2) |
| 0.5 | 49.3 | 50.7 | 1.7(2) |
| 0.7 | 29.3 | 70.7 | 1.6(2) |
| 0.8 | 20 | 80 | 1.7(2) |
| 0.9 | 9.7 | 90.3 | 1.2(1) |
| 1 | - | - | 1.1(1) |

Figure S1 shows a typical graphical output of the Rietveld refinement. It concerns $Ti_{0.3}V_{0.7}H_2$ hydride (*i.e.*, $x = 0.7$) The experimental data is well fitted by modeled diffraction pattern. Only the fcc fluorite-type structure is detected for this synthesis.

---

[*] Corresponding author. E-mail : fermin.cuevas@cnrs.fr
[†] Deceased author

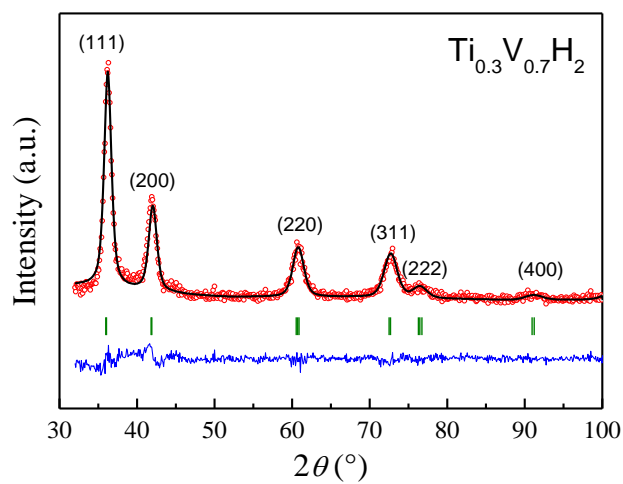

*Figure S1: Graphical output of the Rietveld analysis for $Ti_{1-x}V_xH_y$ hydride for $x = 0.7$. Observed (red dots), calculated (black solid line) and difference curves (blue bottom line) are shown. Vertical bars (/) correspond to Bragg positions ($Cu-k_{\alpha 1,2}$) for fcc-(Ti,V)$H_2$ phase. Peak indexation is given.*

Figure S2 shows the evolution of the specific reversible capacity and coulombic efficiency of $Ti_{0.3}V_{0.7}H_2$ hydride in Li-ion half-cell configuration during galvanostatic cycling. The reversible capacity gradually decreases from 1700 to 50 mAh/g during the first 40 cycles, while the coulombic efficiency increases from 65 to 95%.

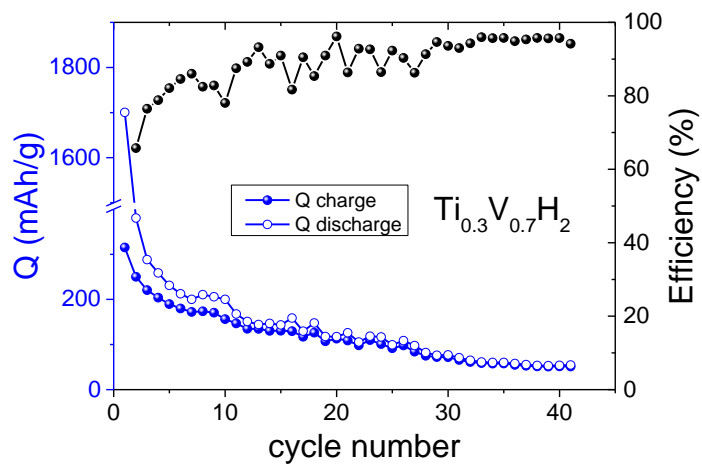

*Figure S2: Evolution of the reversible capacity and coulombic efficiency of 1:1:1 CMC: Carbon Super P: $Ti_{0.3}V_{0.7}H_2$ electrode during galvanostatic cycling.*

Figure S3 displays the evolution of the temperature of the milling vial during the mechanochemistry of vanadium powder under hydrogen gas. The vial temperature increases gradually on milling. As reported previously [1], the vial temperature is ~ 15 °C lower than that of the temperature of the hydrogen gas. In turn, the gas temperature is much lower than the local temperature at the collision impact between milling balls and powder material. The local temperature may attain ~ 200°C as demonstrated in a previous publication [2].

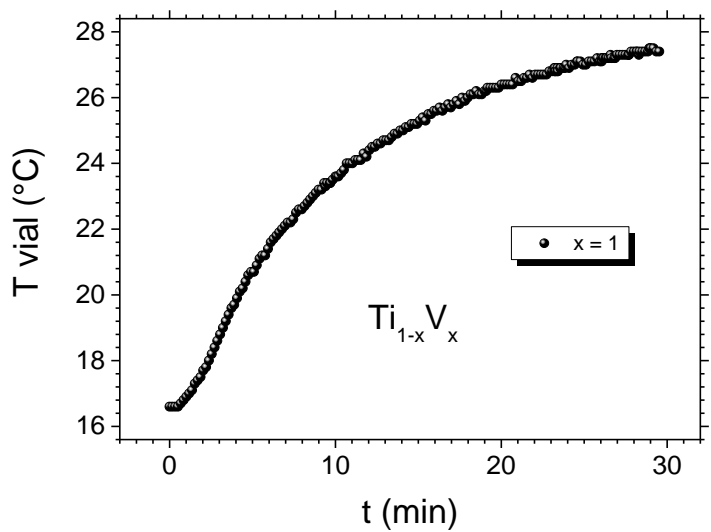

*Figure S3: Evolution of the vial temperature during milling under hydrogen gas of pure vanadium sample.*

Figure S4 shows the differential capacity plots for $Ti_{1-x}V_xH_2$ electrodes within the potential range $0 \leq E \leq 1.2$ V vs. $Li^+/Li$. Besides peak signals related to the conversion reaction between hydrides and lithium, a common peak is detected at 0.9 V and attributed to the SEI formation at the hydride surface.

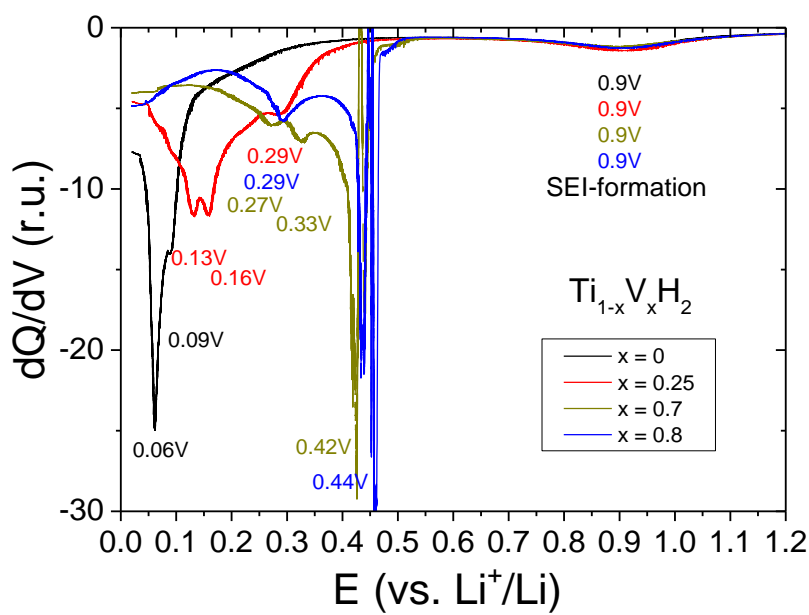

*Figure S4: Differential capacity plots as a function of voltage for selected dihydrides $Ti_{1-x}V_xH_2$ during the first galvanostatic cycle in Li-ion half-cell. Peak potentials are expressly given.*

Figure S5 shows the potential profiles during galvanostatic cycling vs. lithium of a working electrode containing $Ti_{0.3}V_{0.7}H_2$ hydride as active material and a second one without the active material. When the electrode contains $Ti_{0.3}V_{0.7}H_2$ hydride a potential shoulder appears at $E \sim 0.9$ V which is absent for the electrode without active material. This demonstrates that this shoulder is related to a chemical reaction involving $Ti_{0.3}V_{0.7}H_2$ hydride. It has been assigned to SEI formation at the hydride powder. The potential plateau below 0.5 V is attributed to the conversion reaction between the hydride and lithium.

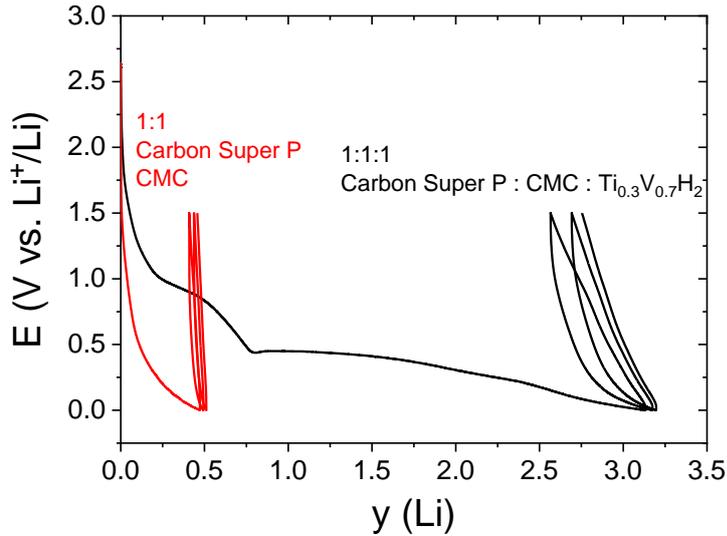

*Figure S5: Comparison between the potential profile of a 1:1:1 CMC: Carbon Super P: $Ti_{0.3}V_{0.7}H_2$ electrode and 1:1 CMC: Carbon Super P during the three first galvanostatic cycles.*

Figure S6 shows the electrochemical and diffraction data recorded during the *operando* XRD experiments. Owing to the differences in electrode formulation, the potential profile in the *in-situ* cell (Figure S6a) differs slightly from that of coin cell, with higher Li-exchange ($y = 0.8$ Li) during SEI formation ($E \geq 0.5$ V vs. Li$^+$/Li) and higher overpotential during the conversion reaction ($E \leq 0.5$ V vs. Li$^+$/Li). The cut-off potential of 0.05 V is reached at $y = 1.82$ Li. Further lithiation up to $y = 2.1$ Li could be achieved after several steps of relaxation for 2 h and lithiation at C/40 regime with same cut-off potential.

Diffraction patters collected for different steps, a-f, of Li reaction are displayed in Figure S6b. Besides diffraction lines from the Be window, fcc-$Ti_{0.3}V_{0.7}H_2$ peaks are clearly detected in all patterns. These peaks gradually decrease in intensity during the electrode lithiation, while broad diffraction peaks gradually appear in the angular domain 38-42° as can be better appreciated in the magnification displayed in Figure S6c. These peaks can be reasonably assigned to the formation of poorly crystallized bct-$Ti_{0.3}V_{0.7}H_z$. The hydrogen content of the bct-phase likely decreases during the conversion reaction since the cell parameters gradually decrease on lithiation (Figure 8b).

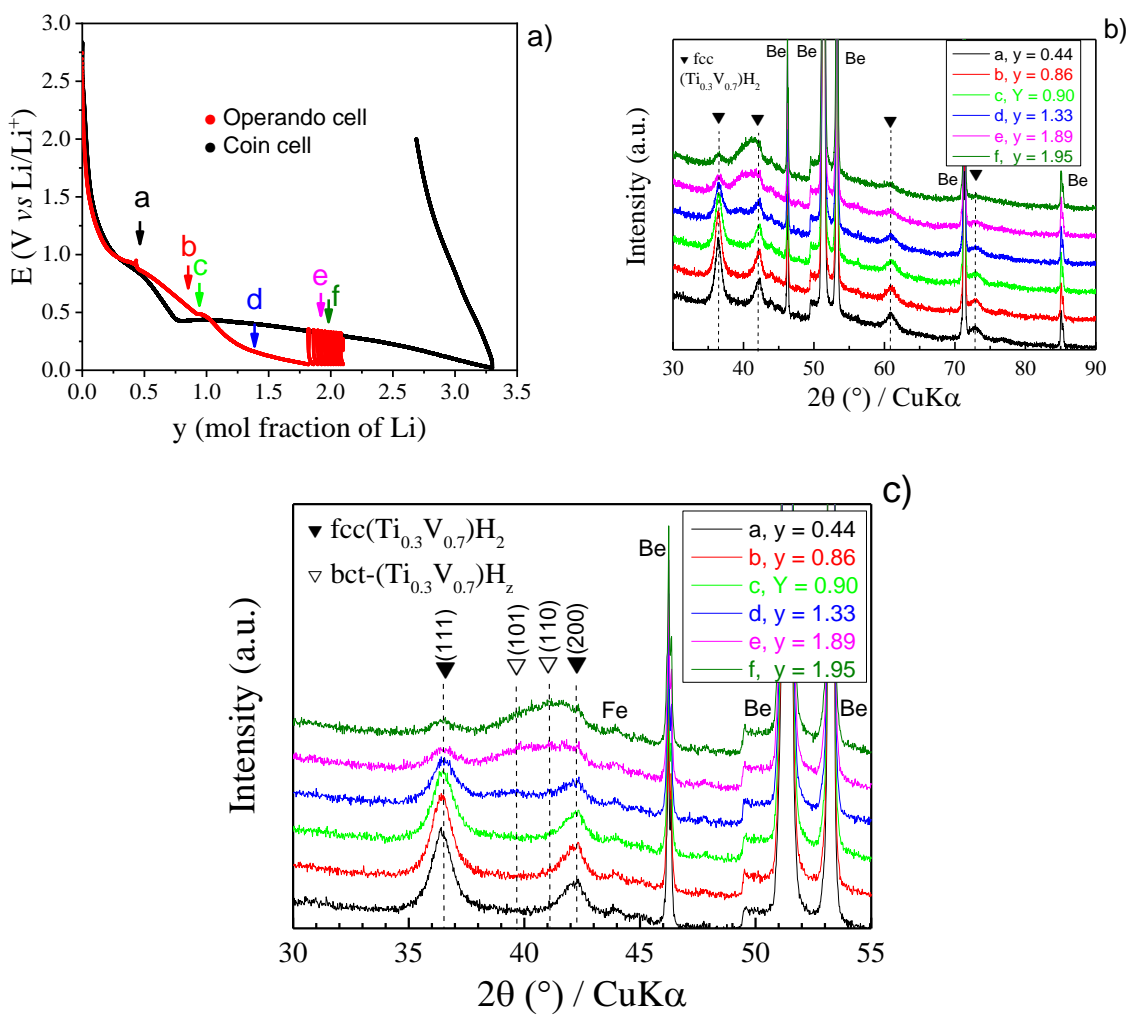

*Figure S6: Electrochemical galvanostatic profiles and diffraction data recorded by operando XRD experiments during lithiation of $Ti_{0.3}V_{0.7}H_2$. a) Comparison of potential profiles collected in coin-type half-cell and in-situ cell at regime C/40, b) Selection of recorded diffraction patterns at different lithiation progress, a-f, marked in figure S6a, c) Magnification of Figure S6b.*

Figure S7 displays the potential profile of the "forced sodiation" with steps of GITT with a current of 25 mA/g. The inset figure shows the XRD pattern of sodiated electrode, the diffraction peaks can be indexed with three phases: Ni (used current collector), the initial hydride $Ti_{0.3}V_{0.7}H_2$ and conversion reaction product NaH. The amount of NaH is evaluated as 46 wt% when considering NaH and $Ti_{0.3}V_{0.7}H_2$ phases. This corresponds to the conversion of 0.95 Na. The formed intermediate hydride or intermetallic compound should be poorly crystalized since no diffraction peaks could be observed for this phase. Another feature of this potential profile is that the potential after each relaxation state equilibrates at ~ 0.2 V vs. $Na^+/Na$

which is close to the calculated potential of the conversion reaction with Na for the compound $Ti_{0.3}V_{0.7}H_2$ (0.26 V vs. $Na^+$/Na).

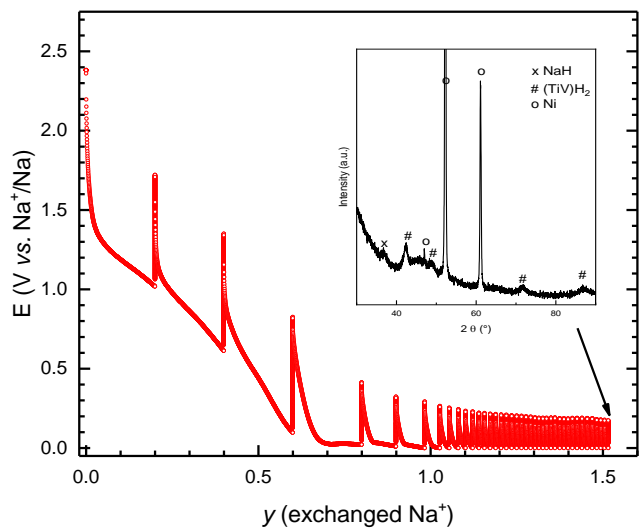

*Figure S7: Forced sodiation potential profile of $Ti_{0.3}V_{0.7}H_2$ with current of 25 mA/g during 4h (4 first steps) and 2h for next steps with a cut-off of 0V vs $Na^+$/Na, along with relaxation periods for 10h at open circuit potential. The XRD pattern of the sodiated electrode at final point.*

Figure S8 shows the Rietveld analysis of the XRD pattern at the end of forced sodiation.

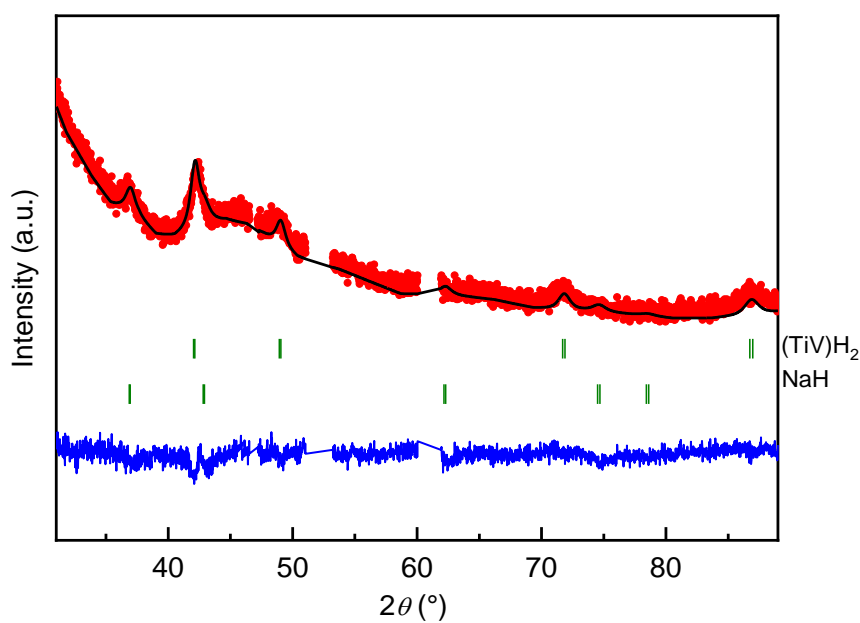

*Figure S8: Rietveld analysis of the XRD (Co-Kα radiation) of the sodiated electrode at the final point shown in Figure S7 ( y = 1.51 Na).*

The same electrode has been dissolved in Suprapur@ HNO$_3$ and pure water. The blank solution was prepared with same quantity of CMC contained the electrode composite. The Na concentration was determined with 4 atomic emission wavelengths. The obtained quantity was calculated to the electrode mass, which corresponds 1.25 mole of NaH per mole of initial Ti$_{0.3}$V$_{0.7}$H$_2$.